\documentclass[final,dvipsnames,5p,times,twocolumn]{elsarticle}
\usepackage[T1]{fontenc}
\usepackage[utf8]{inputenc}
\usepackage{graphicx}
\usepackage{amssymb}
\usepackage{bm}
\usepackage{amsmath}
\usepackage{color}
\usepackage{tikz}

\usepackage{url,hyperref,lineno,microtype}
\hypersetup{
    colorlinks = true,
    citecolor=blue,
    linkcolor=blue,
    urlcolor=blue
}
\usepackage{algorithm}
\usepackage{algorithmic}
\usepackage[flushleft]{threeparttable}
\usepackage{soul}
\usepackage{listings} 
\graphicspath{{fig/}{./fig/}{.}}

\newcommand{\bfm}{\boldsymbol}

\newcommand{\onlinecite}[1]{\hspace{-1 ex} \nocite{#1}\citenum{#1}}
\begin{document}
\begin{frontmatter}
\title{Brute--forcing spin--glass problems with CUDA}
\author[a]{Konrad Jałowiecki\corref{author}}
\author[b]{Marek M. Rams}
\author[b,a]{Bart\l{}omiej Gardas}

\cortext[author] {Corresponding author.\\\textit{E-mail address:} konrad.jalowiecki@smcebi.edu.pl}
\address[a]{Institute of Physics, University of Silesia, Uniwersytecka 4, 40-007 Katowice, Poland}
\address[b]{Jagiellonian University,  Marian Smoluchowski Institute of Physics, \L{}ojasiewicza 11, 30-348 Krak\'ow, Poland}

\begin{abstract}
We demonstrate how to compute the low energy spectrum for small ($N\le 50$), but otherwise arbitrary, spin--glass instances using modern Graphics Processing Units or similar heterogeneous architecture. Our algorithm performs an exhaustive (i.e., brute--force) search of all possible configurations to select $S\ll 2^N$ lowest ones together with their corresponding energies.
We mainly focus on the Ising model defined on an arbitrary graph. An open--source implementation based on CUDA Fortran and a suitable Python wrapper are provided. As opposed to heuristic approaches, ours is exact and thus can serve as a references point to benchmark other algorithms and hardware, including quantum and digital annealers. 
Our implementation offers unprecedented speed and efficiency already visible on commodity hardware. At the same time, it can be easily launched on professional, high--end graphics cards virtually at no extra effort. 
As a practical application, we employ it to demonstrate that the recent Matrix Product State based 
algorithm---despite its one-dimensional nature---can still accurately approximate the low energy spectrum of fully connected graphs of size $N$ approaching $50$.
\end{abstract}
\begin{keyword}
CUDA Fortran \sep 
Ising spin--glass \sep 
Quantum annealers \sep
Titan V GPU
\end{keyword}
\end{frontmatter}
\section{Introduction}
With increasing complexity and interconnectivity in the modern world, the ability to solve optimization problems becomes indispensable.
Notwithstanding, these problems are fundamentally hard to resolve as they often require seeking over {\it enormous} spaces of possible solutions~\cite{Aaronson13}. A notable example is the famous spin--glass problem encoded via the Ising model~\cite{harris_phase_2018}, where the low energy spectrum (the ground state in particular) is sought after. The importance of this system is reflected in the fact that many NP-complete~\cite{garey_computers_1979} optimization problems (i.e.~Karp's $21$ problems~\cite{karp}) can be mapped onto its Hamiltonian~\cite{Lucas14}. Furthermore, there is growing hardware support for many spin--glass based models~\cite{yamamoto_coherent_2017,fujitsu,dwave}. These cutting edge technologies, when combined with classical neural networks~\cite{alexnet}, lead to quantum artificial intelligence~\cite{rbm18}. A type of artificial intelligence believed to be powerful enough to simulate many--body quantum systems \emph{efficiently}, which is a holy grail of modern physics~\cite{elsayed_entangled_2018}.

The most promising ideas to overcome mathematical difficulties concerning classical optimization could rely on quantum computers~\cite{Feynman60}. In particular, on quantum annealers such as the D-Wave $2000$Q chip~\cite{Lanting14}. In principle, such machines could solve variate of (hard) optimization problems (almost) ``naturally'' by finding low energy eigenstates~\cite{orus14}. However, current quantum annealers are extremely noisy and thus not powerful enough to tackle large scale optimization challenges~\cite{GardasDeffner18,Gardas17}. In contrast, heuristic approaches, often offering superior performance, can {\it not} typically certify that the solution that has been found is, in fact, optimal~\cite{czartowski18,baccari_verification_2018}. Most heuristic solvers rely on strategies ranging from famous simulated annealing~\cite{cook2018gpu}, branch and bound approaches~\cite{rendl_solving_2008} their chordal extension~\cite{baccari_verification_2018}, various Monte Carlo methods~\cite{Hen17} throughout dynamical systems simulations~\cite{diventro} to tensor network analysis~\cite{OTN}. 

In this work, we focus on yet another class of solvers, namely those that perform exact brute--force search~\cite{heule_science_2017}. The idea is to search the entire Hilbert space exhaustively to find configurations with the lowest energies. Such a search can be performed either in the probability or energy space~\cite{OTN}. For all classical Hamiltonians, where all their terms commute, this is essentially equivalent to the exact diagonalization. However, in contrast to the quantum case, the eigenvalue problem for classical models can be executed truly in parallel. An efficient implementation nonetheless is {\it not} trivial.

Although practical applications of such solvers are limited to small problem sizes ($N \le 50$), they can solve the Ising model that is defined on an arbitrary graph. Moreover, with the exhaustive search, one can easily certify the output. All of these features are crucial for testing, benchmarking, and validating new methods~\cite{Kazuyuki19}, strategies, and paradigms (e.g., memcomputing~\cite{diventro}) for solving classical optimization problems~\cite{Katzgraber18}. It is worth mentioning that brute--force approaches however limited can still serve as a reference point for today's quantum supremacy experiments~\cite{google,ibm}.

Our implementation offers excellent flexibility and portability, as well as the significant efficiency and speed. Our solver can be executed on either CPU (Central Processing Unit) or GPU (Graphics Processing Unit) using Nvidia's CUDA (Compute Unified Device Architecture). The latter architecture is of particular importance due to its massive parallel capabilities~\cite{Januszewski15,GPU2018}. Moreover, we provide a simple Python wrapper that allows users to access both architectures effortlessly~\cite{Isingpy}.

Finally, we employ our solver to test the applicability of a particular tensor network ansatz---based on the Matrix Product State (MPS)---to optimization purposes of a fully connected graph of growing size. For a detailed description of this algorithm, we refer the reader to look at Supplementary Information in Ref.~\cite{OTN}. We have verified that indeed, such an ansatz, despite its inherited one--dimensional structure, can still successfully capture the low energy spectrum for tested graphs up to $N=50$.
This indicates that the MPS ansatz should still perform well also for much larger systems having a dominant quasi--one--dimensional nature. At the same time, sparse connections at long--range do \emph{not} necessary exclude the applicability of the MPS approach. 

\section{Spin-glass problems}
\begin{figure}[t!]
    \centering
\includegraphics[width=0.4\textwidth]{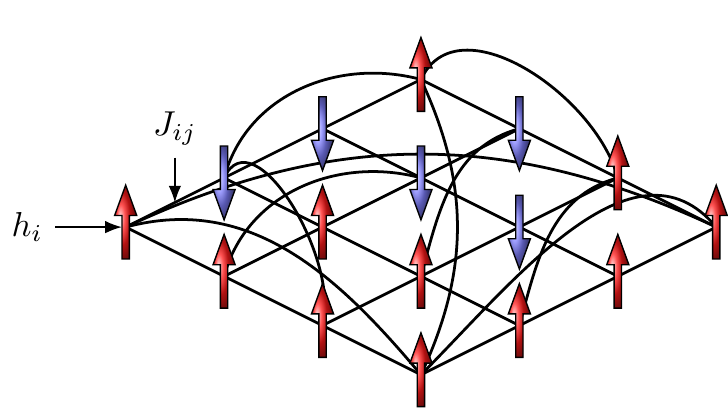}
\caption{An example of the Ising spin glass model~(\ref{eq:ising}). Here, $J_{ij}$ correspond to weights of the edges of the graph and $h_i$ are biases associated with the graph's nodes ($N=16$). Physically, $J_{ij}$ describe the interaction between spins $s_i$, $s_j$ and $h_i$ is the external magnetic field imposed on spin $s_i$.
The picture also demonstrate possible spin encoding, with red arrows indicating assignment of $s_i=+1$ and blue ones indicating assignment of $s_i=-1$ [or $q_i=0$ if QUBO~(\ref{eq:qubo}) is used].
}
    \label{fig:ising}
\end{figure}

In this work, we mainly focus on the Ising Hamiltonian.
However, our approach can easily be extended to include other \emph{classical} spin--glass models~\cite{Wu82,Cai18}.
To begin with, consider a simple undirected graph with $N$ nodes (i.e., vertices) as the one drawn in Fig.~\ref{fig:ising}. We assign a unique spin variable, $s_i\pm 1$ (blue and red arrows), to each node. Adjacent nodes labeled as $i$, $j$ are coupled via interaction strength $J_{ij}$, which may be viewed as a weight of the edge connecting those two nodes. Additionally, for every spin, we associate a local magnetic field (bias) $h_i$ interacting with it.
Then the energy of such a system of spins is defined as 
\begin{align}
\label{eq:ising}
    H(\boldsymbol{s}) = - \sum_{\langle i, j \rangle} J_{ij} s_i s_j - \sum_{i=1}^N h_i s_i,
\end{align}
where $\boldsymbol{s}:=(s_1, \ldots, s_L)$.
The first sum runs over all adjacent sites, which we denote here as $\langle i, j\rangle$. 

In many practical applications, one is typically interested in finding a particular spin configuration, say $\boldsymbol{s}_0$, for which $H(\boldsymbol{s}_0)$ in Eq.~\eqref{eq:ising} admits its minimum value. Such configuration is called the \emph{ground} state. Naturally, states with energies above the ground state energy are called \emph{excited} states. 
Finding the low energy spectrum (consisting of the ground state energy and a number $S\ll 2^N$ of excited states) of the Ising model~(\ref{eq:ising}) can also be formulated as a Quadratic Unconstrained Optimization Problem (QUBO). Namely,
\begin{equation}
\label{eq:qubo}
F(\boldsymbol{q}) = -\sum_{\langle i,j \rangle} a_{ij} q_i q_j - \sum_{i=1}^N b_i q_i,  
\end{equation}
where $\boldsymbol{q} = (\boldsymbol{s} + 1) / 2$ are \emph{binary} variables whereas
\begin{equation}
\label{eq:toQUBO}
a_{ij}= 4J_{ij},
\quad 
b_i= 2h_i - 2 \sum_{\langle i, j \rangle} J_{ij}.
\end{equation}
The energy offset reads $H(\boldsymbol{s})-F(\boldsymbol{q}) = \sum_{i=1}^N h_i - \sum_{\langle i, j \rangle} J_{ij}$. Note, if a given $q_i$ vanishes so does any product $q_iq_j$. Therefore, QUBO formulation~(\ref{eq:qubo}) {\it effectively} reduces the number of multiplications almost by half in comparison to Eq.~(\ref{eq:ising}).

\begin{figure}
    \centering
    \includegraphics[width=0.5\textwidth]{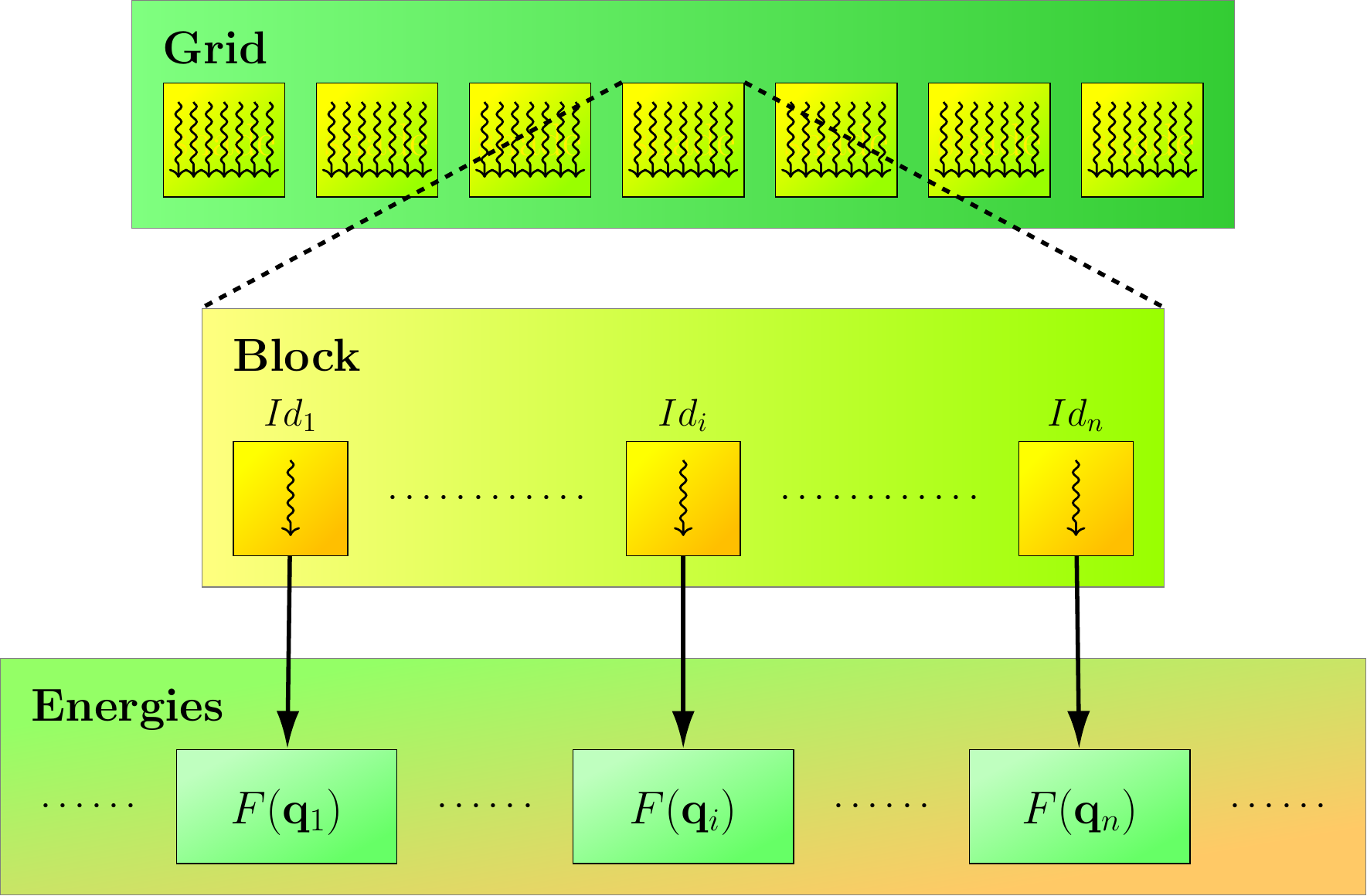}
    \caption{Scheduling of the energy computation on the GPU. A CUDA program is executed by threads that are organized by blocks. Both the grid and blocks can form one, two or three dimensional structures. Our implementation uses a one dimensional grid structure, where the global thread index, $Id_i$, is converted into a state $\boldsymbol{q}$ with mapping $Id_i=(\boldsymbol{q})_2$, cf. Eq.~(\ref{eq:coding}). Next, each thread in each block computes its own energy, $F(\boldsymbol{q})$, according to Eq.~\eqref{eq:toQUBO}. To fit into, often limited, GPU memory the computation is executed in carefully tailored chunks, cf. Eq.~(\ref{eq:codingEx}).
    }
    \label{fig:my_label}
\end{figure}

Despite its straightforward formulation, the problem of solving spin--glass instances can \emph{not} be easily tackled using a brute force approach even for a modest number of spin variables. This is since the number of possible spin assignments grows exponentially with the number of nodes in the graph. For instance, when $N=40$, the number of possible states is greater than the number of bits in a $32$GB memory chip. Already when $N=64$, the size of the search space is greater than the estimated age of the Universe in seconds~\cite{planck_2015}. In fact, the problem of finding the ground state of the Ising model defined on an arbitrary graph is long known to be NP--hard~\cite{barahona_computational_1982}. This means, in particular, that even verifying if a given configuration minimizes the cost function~(\ref{eq:ising}) is difficult. 
\section{Description of the algorithm}
\begin{figure*}[t!]
\label{fig:results}
\centering
\includegraphics[width=\textwidth]{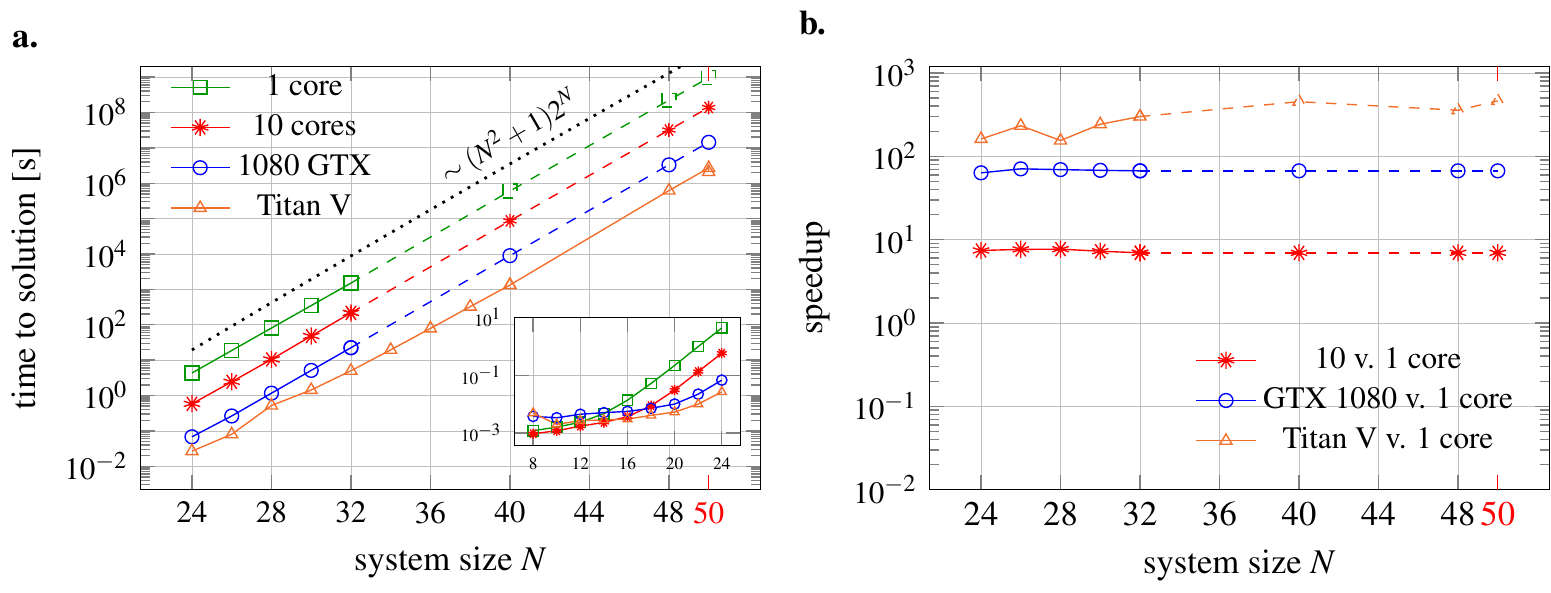}
\caption{{\bf a.} Time to solution obtained by our solver for various system sizes $N$ (the inset captures small systems). {\bf b.} Calculated speedup in comparison to a single CPU core. The speedup is obtained as a ratio of respective execution times.
The algorithm computes $S\ll 2^N$ low energy states in a single run (here $S=10^2$). The problem instances ($J_{ij}$, $h_i$), on a fully connected graph, were randomly generated.
The solid lines show actual measurements: $100$ repetitions for each $N$ except for $N=48$ and $N=50$ for which time to solution was calculated only once. The dashed lines represent \emph{experimentally} estimated values for larger system sizes. The estimate is based on the time necessary to process a single chunk of data [of size of size $M=2^{29}$ (CPU), and $M=2^{27}$ (GTX 1080)]. This estimation is consistent with the scaling~(\ref{eq:cmplx}), which is depicted as a black dotted line.
For $N \ge 24$ the overhead of parallel computations starts playing less important role and the execution time becomes linear in the state space size.
}
\end{figure*}
A general idea underlying this work is to perform an exhaustive search over the whole state space, taking advantage of massive parallel capabilities of modern GPUs. 
This requires an efficient strategy to encoding all states, $\boldsymbol{q}=(q_1, q_2,\dots, q_L)$, on a GPU. A naive approach would required storing an array of $N$ integers, $q_i=0,1$, for each state $\boldsymbol{q}$. However, this would also lead to excessive use of memory and render this approach inefficient. As an optimal strategy, one should try to reuse information already stored in the GPU memory. 
Therefore, in our algorithm, we take advantage of the following correspondence
\begin{equation}
\label{eq:coding}
\text{GPU thread index} = (\boldsymbol{q})_2,  
\end{equation}
where $k=(\boldsymbol{q})_2$ denotes the binary representation of an integer $k$. For instance, when
there is $N=8$ spins, one may associate

\begin{equation}
\label{eq:codingEx}
\text{thread index}\; \# 13 = (00001101)_2,  
\end{equation}
%
Theoretically, this strategy allows one to store $M=2^{64}\sim 10^{19}$ states with no extra cost, limiting the system size to $N=64$ spins.  Nonetheless, this is more than the current architecture, based on the von Neumann paradigm of computation, which can process in a reasonable time~\cite{Backus_Bottleneck}. Indeed, we estimated that optimal search among $2^{64}$ states to extract the low energy spectrum consisting of $S=10^2$ of them would take $821$ \emph{years} on an efficient Titan V GPU~\cite{titanv}. 
In comparison, systems of sizes $N=32$, $49$, $50$ can be solved within $5$ seconds, $12$ and $24$ days, respectively. A detailed benchmark is presented in Sec.~\ref{sec:benchmark}.

One should stress that the fastest (as of $2018$) supercomputer in the world---\href{https://www.olcf.ornl.gov/olcf-resources/compute-systems/summit/}{Summit}---is equipped with $27648>2^{14}$ Nvidia Tesla V$100$ GPUs~\cite{summit}. Therefore, ``only'' $2^{14}$ of them (processing chunks of size $2^{50}$ each, simultaneously) should be able to reduce the number of $821$ years (for a single GPU tackling $N=64$) substantially. Perhaps, maybe even down to a couple of months. Nevertheless, \emph{a priori}, it is hard to estimate the exact numbers due to various communication bottlenecks. This interesting open problem is, however, beyond the scope of the current work. 

In theory, one could first compute all $M=2^N$ energies in parallel and only then select $S\ll 2^N$ lowest ones (and the corresponding states if needed). However, even with an efficient storage strategy, this approach quickly becomes impractical for large systems. It requires an exponentially increasing storage space to encode possible solutions. To overcome this problem, one could iterate over the solution space in manageable chunks, each time extracting the desired number of states [e.g., with the bucket select algorithm~\cite{alabi_fast_2012}, for which the mean execution time scales linearly with the size of the input vector]. Sorting the energies is executed only in the final step. Since GPU threads and blocks are labeled in the same way for every chunk, an offset is required to correctly enumerate all states, i.e.,
\begin{equation}
\label{eq:coding2}
\text{GPU thread index} + \text{offset} = (\boldsymbol{q})_2,    
\end{equation}
Note, the energy calculations are independent and thus can be performed in parallel. The overall parallel speedup is limited by the serial part (Amdahl's law~\cite{hill_amdahls_2008}) consisting of the lowest energy states extraction and merging all local information into the global record. 
Algorithm~\ref{alg:search} in the below listing summarizes the underlying structure of our solver.
\begin{algorithm}
\caption{Searching $S\ll 2^N$ configurations (i.e. states) with the lowest energies defined in Eq.~(\ref{eq:ising}). The adjacency matrix, $J_{ij}$, and local magnetic fields, $h_i$, are provided.}
\label{alg:search}
\begin{algorithmic}
\STATE k $\leftarrow$ chunk\_size\_exponent

\FOR{$i=1$ to $2^{N-k}$}
   \FOR{$j=1$ to $2^{k}$}
       \STATE state\_code $\leftarrow j + (i - 1) \cdot 2^{k}$
       \STATE energies[$j$] $\leftarrow$ energy(graph, state\_code)
       \STATE states[$j$] $\leftarrow$ state\_code
   \ENDFOR
   \STATE select\_lowest(energies, states, num\_st)
   \IF{$i == 1$}
      \STATE low\_en[$1\colon\mbox{num\_st}$] $\leftarrow $energies[$1\colon\mbox{num\_st}$]
      \STATE low\_st[$1\colon\mbox{num\_st}$] $\leftarrow $states[$1\colon\mbox{num\_st}$]
   \ELSE
      \STATE low\_en[$\mbox{num\_st}+1\colon2 \cdot\mbox{num\_st}$] \\ $\qquad \leftarrow $energies[$1\colon\mbox{num\_st}$]
      \STATE low\_st[$\mbox{num\_st}+1\colon2 \cdot\mbox{num\_st}$] $\leftarrow $states[$1\colon\mbox{num\_st}$]
      \STATE select\_lowest(low\_en, low\_st, num\_st)
   \ENDIF
\ENDFOR

sort\_by\_key(low\_en, low\_st, num\_st)
\end{algorithmic}
\end{algorithm}
\section{Implementation details}
\subsection{Languages and technologies employed}
The core components of our implementation has been written in modern Fortran $95/2003$~\cite{chapman07}, which we have chosen for its flexibility~\cite{rossi_pymiedap_2018}, extensive support for linear algebra~\cite{wang_intel_2014}, performance~\cite{julia_benchmark} and native support for CUDA technology~\cite{cuda_fortran}. 
To make our code easier to use, we have wrapped it in a Python package using the f2py~\cite{f2py} utility and numpy's fork of distutils package~\cite{numpy_distutils}.
Whereas Fortran is widely used mostly for numerical simulations~\cite{Wall12}, Python is one of the 
most popular general--purpose programming language~\cite{so_survey}.

We have also incorporated the fast $k$--selection algorithm for GPU~\cite{alabi_fast_2012}, and the Thrust library~\cite{thrust} into our solver for its parallel implementation of many standard methods such as finding the minimum and maximum of an array or partitioning thereof. The Thrust library is utilized both for the GPU implementation and for the pure CPU implementation with the OMP backend. In order to take advantage of various Thrust's functions, we have written several small C++ modules to bind them into the Fortran code. The source code of the entire package, together with the comprehensive documentation, can be found on GitHub~\cite{Isingpy}.

The Python wrapper allows one to execute the algorithm both on the CPU and GPU. It was designed with simplicity in mind, and as such, its primary usage does \emph{not} require any specialized knowledge. A basic understanding of the underlying optimization problem is enough, cf. the listing below. In particular, only the system definition (i.e., the graph or adjacency matrix) and the desired number of states needs to be provided by the user. Nevertheless, other parameters, including the chunk sizes, can also be passed to the wrapper. 
\begin{figure*}
    \centering
    \includegraphics[width=\textwidth]{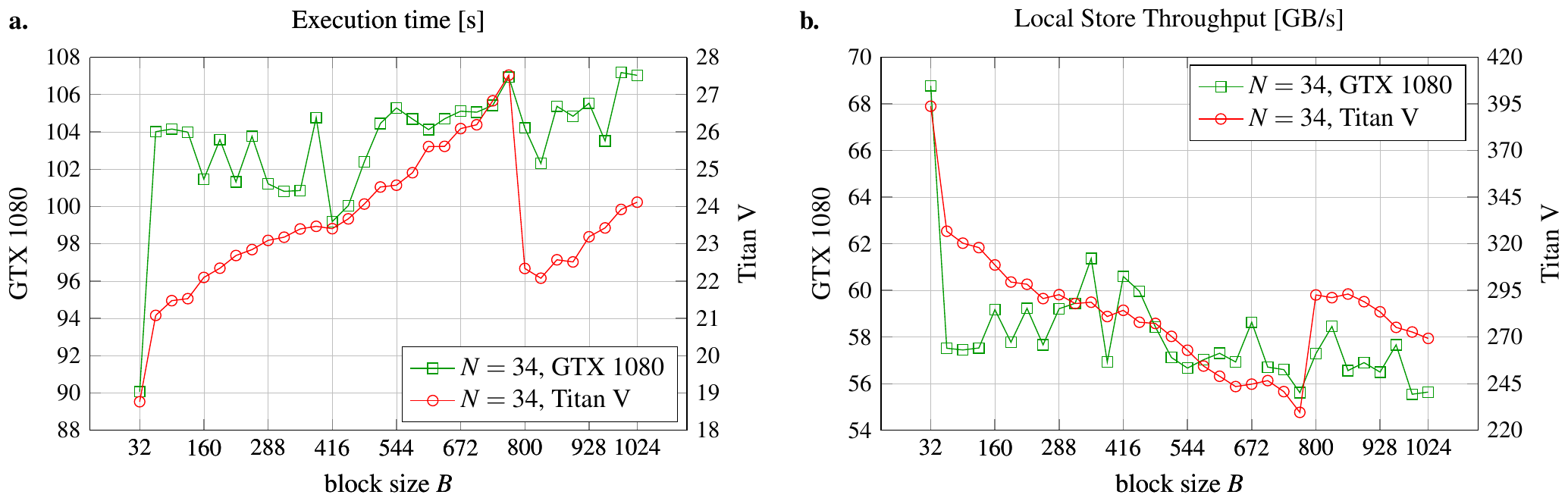}
\caption{    
    {\bf a.} The execution time (in seconds) of the proposed algorithm versus the CUDA block size, $B$, for a given system size, here $N=34$. A clear (global) minimum is visible for $B=2^5=32$, which is exactly the number of threads in the warp, i.e., the warp size. This behavior anticorrelates with the local storage throughput shown in {\bf b.} 
    The optimal value of $B=32$ was used for all benchmarks presented in this work.}
    \label{fig:B}
\end{figure*}

\begin{lstlisting}[language=Python, caption=Simple example of how to use the \texttt{ising} module]
# import package
from ising import search

# adjacency matrix (problem definition)
graph = {(1,1):-1, (1,2):-0.2}
# solve the Ising model
solution = search(graph, num_states=4)

# shows the states and energies found
print(solution.energies)
print(list(solution.states))
\end{lstlisting}
On virtually all Linux platforms, it is possible to install the very basic version (i.e., with no GPU support) of our solver directly from the Python Package Index, by issuing
\begin{equation}
\texttt{pip install ising}
\end{equation}
where \texttt{ising} is the name of the package. However, to assure full compatibility with modern GPUs, CUDA requires a custom build from source which can be initiated via
\begin{equation}
\texttt{python install.py {-}{-}usecuda}
\end{equation}
from the package source directory.
For more details regarding custom installation, including CUDA and various Fortran compilers, we refer the reader to documentation~\cite{ising_doc}. Note, only PGI and IBM XLF support CUDA Fortran. Our package has only been tested using the former.
\subsection{GPU execution scheduling}

Programming GPUs often pose a nontrivial endeavor. Among many challenges, one has to design the grid on which kernels are launched~\cite{sanders_cuda_2010}. 
We have tested various grid/blocks launching configurations for the energy computing kernel and have obtained the best results with grids consisting of $G=2^{g-5}$ ($g$ being the current chunk size) blocks of size $B=2^5$ each. This particular value maximizes the Local Store Throughput (LST)---reported by the NVIDIA Profiler, \emph{nvprof}---and thus also the execution time of the energy computing kernel, cf. Fig.~\ref{fig:B}. 

Note also that the Titan V is roughly $6$ times faster than GTX $1080$, which is exactly the ratio between LSTs for these two devices. Such apparent correlation validates, to some extent, the LST as a proper metric (in comparison to a typical Throughput used to benchmark GPUs) to asses the performance of our solver.

We would like to stress, however, that the optimal launching configuration we have used in the present studies may need further (experimental) adjustment depending on the hardware, and possibly the problem size.

\subsection{Complexity analysis}
\label{sec:cplx}
Our algorithm performs an exhaustive search over the entire, exponentially large, state space in predefined chunks to find $S$ lowest states (cf. Algorithm~\ref{alg:search}). Thus, unavoidably its time complexity has to be at least exponential in the system size $N$. Computing the energy~(\ref{eq:qubo}) for a single state, $\boldsymbol{q}$, requires $\mathcal{O}(N^2)$ operations. The selection procedure executed on a data chunk of size $2^k$, however, requires $\mathcal{O}(2^k)$ comparisons resulting in $\mathcal{O}[2^k(N^2+1)]$ operations. Finally, taking into account the total number of chunks, $2^{N-k}$, and adding complexity of the final sorting procedure, $\mathcal{O}[S\log(S)]$, results in total complexity being 
\begin{equation}
\label{eq:cmplx}
\mathcal{O}[2^N(N^2+1) + S\log(S)] = \mathcal{O}[2^N(N^2+1)].
\end{equation}
Therefore, essentially the solver's complexity behaves as $\mathcal{O}(2^N)$ which we also demonstrate experimentally in Sec.~\ref{sec:benchmark} (cf. Fig.~\ref{fig:results}). 
\begin{figure*}[ht]
    \centering
    \includegraphics[width=\textwidth]{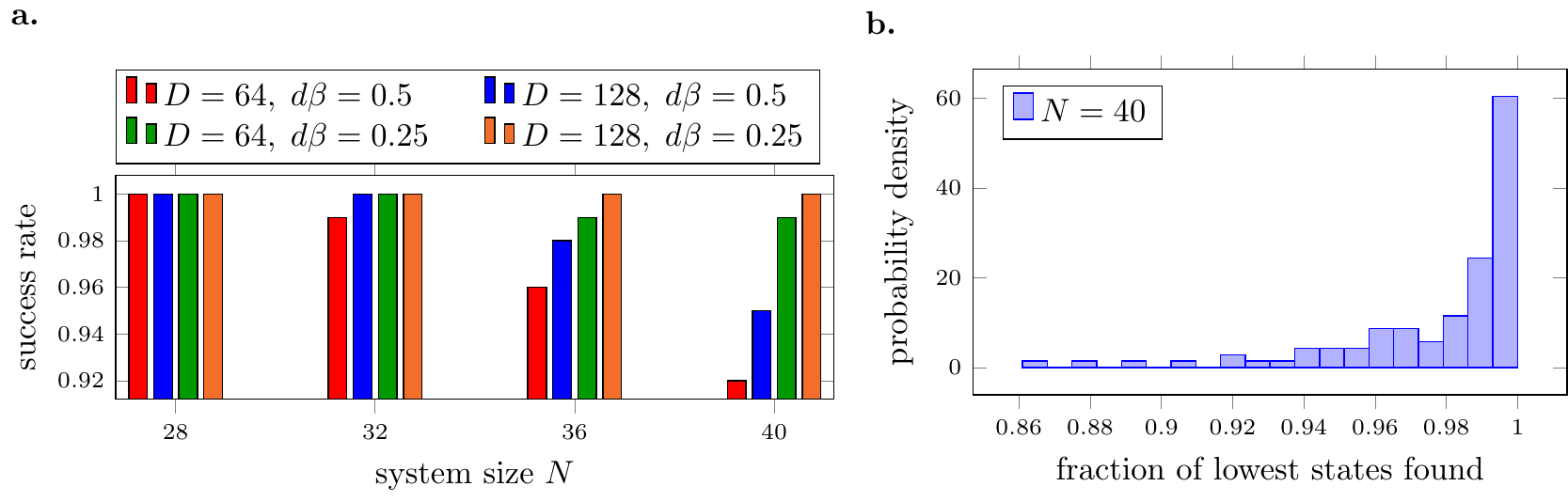}
    \caption{
    Verification of the Matrix Product State (MPS) based algorithm introduced in Ref.~\cite{OTN}.
    {\bf a.} Success rate is defined as the fraction of cases for which the MPS algorithm was able to find the ground state. All spin-glass instances were generated randomly on a fully connected graph of size $N$. Parameter $D$ is the bond dimensions characterizing MPS tensors, and $d\beta$ denotes the increment of the inverse temperature, see the main text. {\bf b.} Normalized histogram showing the percent of instances for which the MPS based algorithm was able to find a given number of configurations out of $S=1000$ lowest ones.
    Here, we use $D=128$, $d\beta=0.25$ and in all panels $\beta=1$. 
    }
    \label{fig:stats}
\end{figure*}
As one can see, the GPU implementation takes $30$ seconds (GeForce $1080$) and $5$ seconds (Titan V) on average to solve the Ising problems with $N=32$ spins. The same problem requires about $1500$ seconds on average on a single CPU core. For GPU, the differences in solution times between single and double precision are close to $10\%$ and are not reported on Fig.~\ref{fig:results}.

\section{Benchmarks}
\label{sec:benchmark}

\subsection{GPU vs CPU comparison}

We have tested our algorithm on the following hardware:
\begin{itemize}
\item CPU: \href{https://ark.intel.com/products/94456/Intel-Core-i7-6950X-Processor-Extreme-Edition-25M-Cache-up-to-3-50-GHz-}{$10$ Cores ${\rm Intel}^{\rm R}$ ${\rm Core}^{\rm TM}$ i7-$6950$X};
\item GPU(1): \href{https://www.nvidia.com/en-us/geforce/products/10series/geforce-gtx-1080}{Nvidia GeForce GTX $1080$, $8$GB GDDR$5$ global memory, $2560$ CUDA Cores};
\item  GPU(2): \href{https://www.nvidia.com/en-us/titan/titan-v/}{Nvidia Titan V, $12$GB HBM$2$ global memory, $5120$ CUDA Cores}.
\end{itemize}

For benchmarking purposes, we have executed our algorithm on a fully connected, $K=100$ randomly generated (cf. Ref.~\cite{Hen16,Firas18}), problem instances for systems up to $N=50$ (on Titan V). For each instance, we have calculated the low energy spectrum consisting of $S=10^2$ states in a single run. Typical results obtained with a high-end CPU (i7-$6950$X) and both a mid-class (GeForce $1080$) and professional (Titan V) GPU are depicted in Fig.~\ref{fig:results}. We have also  estimated time to solution \emph{experimentally}, for larger systems (up to $N=50$ spins) for which the low energy spectrum can be obtained in a reasonable time (i.e., one month) on Titan V. The estimate is based on the average time required to process a single chunk of data [of size $M=2^{29}$ (CPU), and $M=2^{27}$ (GTX $1080$)]. Our measurements are consistent with the complexity analysis discussed in Sec.~\ref{sec:cplx}.

\subsection{Validation of MPS algorithm}
To demonstrate the capabilities of our solver, we employ it to benchmark a more sophisticated, {\it heuristic}, approach based on a Matrix Product States (MPS) technique (see Supplementary Material of Ref.~[\onlinecite{OTN}] for details). Here, we are not interested in time to solution, but instead, we would like to investigate the accuracy of the latter. Heuristic algorithms can often solve large systems ($N\gg 50$). However, they cannot certify solutions.

With the MPS based algorithm one aims at approximating the Boltzmann distribution,
\begin{equation}
e^{-\beta H(\bfm s)/2} \approx A^{s_1} A^{s_2} \ldots A^{s_L} = |\Psi(\beta) \rangle,
\label{eq:MPS_approx}
\end{equation}
for a sufficiently large inverse temperature, $\beta$, where each $A^{s_i}$ ($i = 1,2\ldots,L$) is matrix of limited dimensions $\le D$ (refereed to as the \emph{bond dimensions}). The above approximation is usual depicted---using a network of tensors---as 
\begin{equation*}
\begin{aligned}
\includegraphics[width=0.95\columnwidth]{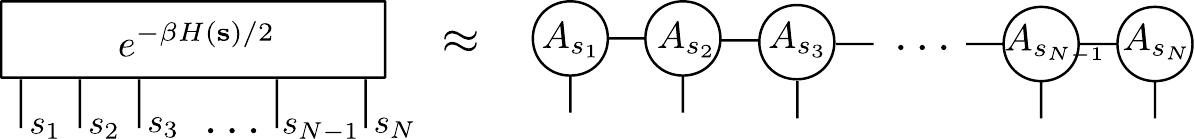}
\end{aligned}.
\end{equation*}
At each bond, one splits the system into two--halves.  The {\it exact} decomposition would require the bond dimension $D$ to grow exponentially with the number of spins in one half, interacting with spins in the second half (and arbitrary numerical precision). The limited bond dimension $D$ reflects on the amount of entanglement/correlations (related to a given bipartition), which can be stored in a ``quantum system'' decomposed as MPS~\cite{Wall18}---here we understand a ``quantum system'' as a superposition over all possible classical spin configurations. 
Having the approximation in Eq.~\eqref{eq:MPS_approx} in the form of MPS, we can efficiently calculate any marginal and conditional probability (at the inverse temperature $\beta$) described by $|\Psi(\beta) \rangle$, and then systematically search for the most probable classical configurations (i.e., the ones with the smallest energies) using branch and bound strategy---building the most probable spin configurations one spin at the time.

Finally, to perform the search one needs to find $|\Psi(\beta)\rangle$, which is obtained by starting from $\beta=0$---for which the MPS decomposition $|\Psi(\beta=0) \rangle$ is trivial---and then subsequently simulating the imaginary time evolution (i.e., the annealing).
To that end, we apply the sequence of operators,
\begin{equation}
\label{eq:gate}
U_i(d\beta) = e^{-d\beta s_i (\sum_{j>i}J_{ij}  s_j+h_i)/2},
\end{equation}
which amount to $\prod_{i=1}^N U_i(d\beta) = e^{-d \beta H(\bfm s)/2}$.
Applying each gate~(\ref{eq:gate}) results in doubling of the affected bond dimensions. Moreover, applying all such operators  would result in uncontrollable, exponential growth of the MPS matrices. However, the one--dimensional (and loop--free) structure of the MPS ansatz allows one to systematically, at each step, find its approximation, which effectively compresses the information and maintains the bond dimensions limited to $D$. The whole procedure can be graphically depicted as
\begin{equation*}
\begin{aligned}
\includegraphics[width=0.95\columnwidth]{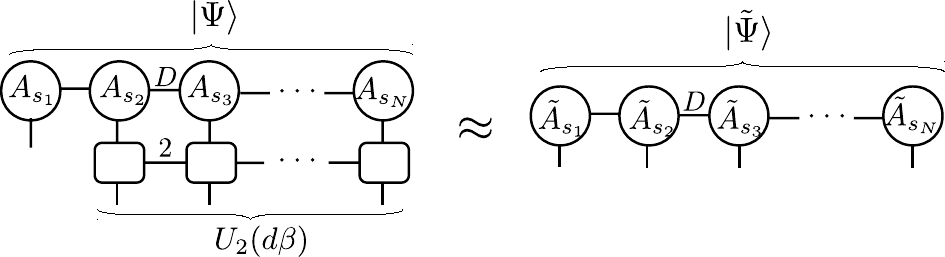}
\end{aligned}.
\end{equation*}
While all the applied operators $U_i(d\beta)$ formally commute (independent of $d\beta$), due to the finite numerical precision and finite $D$, it is relevant to reach the final inverse temperature, $\beta$, gradually in a couple of consecutive steps, each with smaller $d\beta$. Otherwise, for larger $d\beta$, $U_i(d\beta)$ effectively act as projectors trapping the system in a local minima.

The question then becomes how well the MPS ansatz, which by construction is one--dimensional, is able to encode the structure of low energy spectrum for fully--connected graphs. In general, the bigger the system, the higher $D$ necessary to faithfully capture the structure of low energy spectrum.
We observe that already moderate $D$ of $128$ is enough to find all ground states for $100$ considered instances, see Fig.~\ref{fig:stats}a. The inverse temperature $\beta=1$ is large enough to sufficiently zoom--in on the low energy states. At the same time, the importance of small enough time--step (here $d\beta =0.25$) is clearly visible.
It is also enough to recover most of the $1000$ configurations with lowest energies for those instances, see Fig.~\ref{fig:stats}b for $N=40$. Note that the {\it exact} MPS decomposition would require the bond dimension of $2^{N/2} = 2^{20}$. This demonstrate the magnitude of the compression of the relevant information encoded in MPS.

\begin{figure}[ht]
    \centering
    \includegraphics[width=\columnwidth]{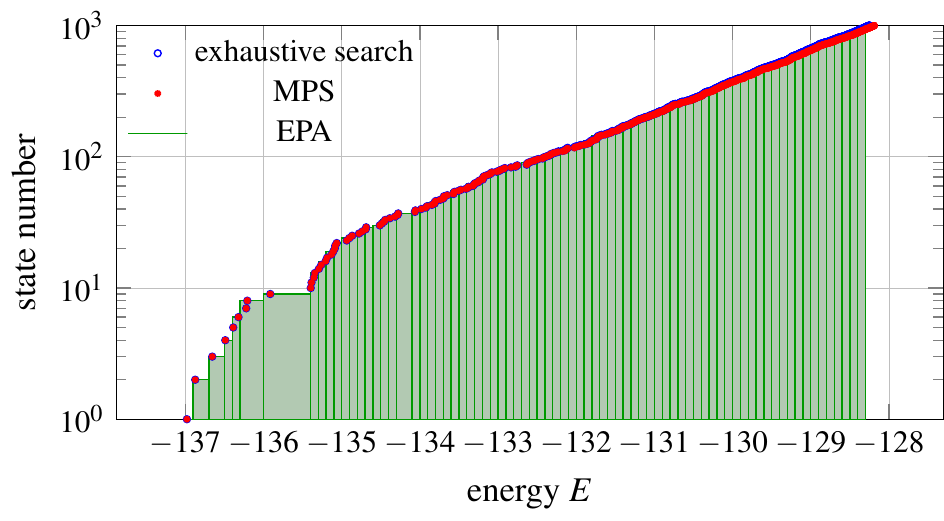}
    \caption{Low energy spectrum of the Ising model~(\ref{eq:ising}) obtained with different algorithms and randomly generated instances of size $N=50$. Here, the bond dimension for the MPS based algorithm $D=128$ and the increment of the inverse temperature $d\beta=0.125$ (cf. the main text or Ref.~\cite{OTN} for more details). We also depicted the approximated solutions obtained with the recent Monte Carlo based algorithm (Entropic Population Annealing), introduced in Ref.~\cite{Barash_2019}. The data was provided by the authors of this paper.
    }
    \label{fig:MPS_PEPS}
\end{figure}

A typical lowest energy spectrum for $N=50$ spins, and consisting of $S=10^3$ states, is shown in Fig.~\ref{fig:MPS_PEPS}. 
Therein, we have also incorporated an approximated low--energy spectrum obtained with the recent Monte Carlo based algorithm, introduced in~\cite{Barash_2019}, which can determine the density of states. The numerical data was provided to us by the authors of that paper.

\section{Summary}
We have demonstrated how to perform an exhaustive (brute--force) search in the solution space of the Ising spin--glass model~\cite{harris_phase_2018} utilizing modern Graphics Processing Units~\cite{pharr_gpu_2005}. Our algorithm can also be adapted for different heterogeneous architectures (e.g., Xeon Phi~\cite{surmin_particle--cell_2016}). The Hamiltonian of this particular model encodes a variety of important optimization problems~\cite{Lucas14}. Moreover, this model has also been realized experimentally as a commercially available D-Wave quantum annealer~\cite{dwave}.

Our implementation with CUDA Fortran~\cite{cuda_fortran} offers unprecedented speed and efficiency already visible on commodity hardware (e.g., GeForce $1080$). Furthermore, it can be easily tuned for professional GPUs such as Titan V~\cite{titanv} virtually at no extra effort. To give an example, our algorithm, when tailored for the latter GPU, can extract the low energy spectrum (consisting of $N=10^3$ states) in roughly $5$ seconds for the spin system admitting $M=2^{32}\sim 10^9$ different configurations. In comparison, a single CPU core takes (on average) $25$ minutes to finish the same task (cf. Sec.~\ref{sec:benchmark} for detailed benchmark). 

Admittedly, practical applications of brute--force algorithms are constrained to small problem sizes ($N \le 50$). However, they can not only solve the spin--glass problems for arbitrary topologies and instances but also certify solutions~\cite{heule_science_2017,czartowski18,baccari_verification_2018}. 
These two features are crucial for developing and validating new methods and strategies for solving classical optimization problems~\cite{Katzgraber18}. We have explicitly exemplified this point by comparing our algorithm to a  sophisticated recent Ising solver based on tensor network techniques~\cite{OTN}.
In particular, we have demonstrated that despite its one--dimensional nature, the Matrix Product State ansatz is still able to approximate well the relevant part of the Boltzmann distribution for a fully connected graph of $N\le 50$. Therefore, this suggests that the MPS algorithm should be superior for all problems having a dominant quasi--one--dimensional nature that allows for sparse connections to span the full problem.

Finally, to benefit the community, we have made our code publicly available as an open--source project~\cite{Isingpy}. Moreover, for those users who lack technical knowledge of Fortran or CUDA, we have provided an easy to install and use Python wrapper~\cite{Isingpy}.

\section*{Acknowledgments}
We appreciate fruitful discussions with Andrzej Ptok, Jerzy Dajka, and Piotr Gawron and Masoud Mohseni. We thank Pawe\l{} Wasiak for his valuable remarks regarding solver's documentation. We gratefully acknowledge the support of NVIDIA Corporation with the donation of the Titan V GPU used for this research. This work was supported by National Science Center (NCN, Poland) under projects 2015/19/B/ST2/02856 (KJ) 2016/20/S/ST2/00152 (BG) and NCN together with European Union through QuantERA ERA NET program 2017/25/Z/ST2/03028  (MMR). MMR acknowledges receiving Google Faculty Research Award 2017.

\bibliographystyle{apsrev4-1_nature}
\bibliography{references}
\end{document}